%% file: PRL_publish.tex
\documentclass[prl,twocolumn,showpacs,amsmath,amssymb]{revtex4-1}
\usepackage{graphicx}
\usepackage{epsfig}
\usepackage{overpic}
\usepackage{dcolumn}
\usepackage{bm}
\usepackage{color}
\usepackage{lineno}
\usepackage{xspace}
\usepackage{verbatim}
\usepackage{subfigure}
\usepackage[colorlinks, linkcolor=blue, urlcolor=blue, anchorcolor=blue, citecolor=blue]{hyperref}
\usepackage{booktabs,tabularx}
\usepackage{array}
\usepackage{titlesec}
\usepackage{multirow}
\usepackage{natbib}

\begin{document}
\normalsize
\parskip=5pt plus 1pt minus 1pt

\title{\boldmath Determination of the $\Sigma^{+}$ Timelike Electromagnetic Form Factors}
\input{authorlist_2023-05-14.tex}
\begin{abstract}

Based on data samples collected with the BESIII detector at the BEPCII collider, the process $e^{+}e^{-} \to \Sigma^{+}\bar{\Sigma}^{-}$ is studied at center-of-mass energies $\sqrt{s}$ = 2.3960, 2.6454, and 2.9000~GeV. Using a fully differential angular description of the final state particles, both the relative magnitude and phase information of the $\Sigma^{+}$ electromagnetic form factors in the timelike region are extracted. The relative phase between the electric and magnetic form factors is determined to be $\sin\Delta\Phi$ = -0.67~$\pm$~0.29~(stat)~$\pm$~0.18~(syst) at $\sqrt{s}$ = 2.3960~GeV, $\Delta\Phi$ = 55$^{\circ}$~$\pm$~19$^{\circ}$~(stat) $\pm$~14$^{\circ}$~(syst) at $\sqrt{s}$ = 2.6454~GeV, and 78$^{\circ}$~$\pm$~22$^{\circ}$~(stat) $\pm$~9$^{\circ}$~(syst) at $\sqrt{s}$ = 2.9000~GeV. For the first time, the phase of the hyperon electromagnetic form factors is explored in a wide range of four-momentum transfer. The evolution of the phase along with four-momentum transfer is an important input for understanding its asymptotic behavior and the dynamics of baryons.

\end{abstract}
\maketitle

Hyperons have a very similar quark composition to that of nucleons, except that one or more of the up or down quarks is replaced by strange quarks. Together with the nucleons, they form a spin-1/2 baryon octet under SU(3) symmetry~\cite{GELLMANN1964214,Zweig}. As one of the fundamental physics observables of the baryons, electromagnetic form factors~(EMFFs) provide a valuable perspective for understanding baryon structure~\cite{A1:2010nsl, Ramalho:2012pu, Eichmann:2016yit} by probing internal charge and current distributions~\cite{Hofstadter:1956qs, Brodsky:1974vy, Geng:2008mf, Green:2014xba}. The EMFFs are analytic functions of the four-momentum transfer squared~($q^{2}$), and they can be divided into spacelike~($q^{2}<0$) and timelike~($q^{2}>0$) regions~\cite{Cabibbo:1961sz, Punjabi:2015bba}. The former are often measured using electron-baryon elastic scattering experiments, while the latter use electron-positron annihilation into baryon antibaryon pairs or the reverse reaction. However, owing to the difficulties in producing stable and high-quality hyperon beams, it is challenging to study the EMFFs of hyperons in the spacelike region. Currently, only a few experiments have measured the EMFFs of hyperon in the spacelike region by elastic scattering of hyperon beam off atomic electrons, and the range of $|q^{2}|$ for exploring EMFFs is limited due to kinematic constraints~\cite{SELEX:2001fbx}. On the other hand, hyperons can be readily produced in electron-positron annihilation above their pair production thresholds. Therefore, the hyperon EMFFs are usually studied in the timelike region via $e^{+}e^{-} \to \gamma^{*} \to Y \bar{Y}$, where $Y$ represents a hyperon with spin $1/2$, and these can be related to the spacelike region via dispersion relations~\cite{Pfister}.   

A large number of measurements are available in the literature for the effective form factors~($G_{\rm eff}$) of SU(3) baryons, which are extracted from production cross sections for $e^{+}e^{-} \to \gamma^{*} \to B\bar{B}$ under the assumption of the electric form factor~($|G_{E}|$) equal to magnetic form factor~($|G_{M}|$)~\cite{Bisello:1983at, DM2:1990tut, Antonelli:1998fv, BES:2005lpy, Achasov:2014ncd, BESIII:2015axk, CMD-3:2015fvi, BESIII:2017hyw, BESIII:2019nep, BESIII:2019cuv, BESIII:2019hdp, BESIII:2020ktn, BESIII:2020uqk, BESIII:2021aer, BESIII:2021tbq, BESIII:2021rkn}. Previous measurements also exist for the modulus of EMFF ratios $\vert G_{E}/G_{M} \vert$, which are obtained by analyzing one-dimensional angular distributions~\cite{CMD-3:2015fvi, BESIII:2015axk, BESIII:2019hdp, BESIII:2020uqk}. However, according to the optical theorem, the form factors at the lowest order for the spacelike region are real due to the Hermiticity of the electromagnetic Hamiltonian, while in the timelike region they are complex~\cite{Tomasi-Gustafsson:2005svz, Denig:2012by}. Thus, a complete knowledge of EMFFs includes the relative phase $\Delta\Phi$ between electric and magnetic form factors, $G_{E}$ and $G_{M}$. Since a nonzero $\Delta\Phi$ ensures a transverse polarization for the produced baryons~\cite{Dubnickova:1992ii},
$\Delta\Phi$ can be extracted from the polarization. The transverse hyperon polarization is self-analyzed in their weak decays, while the polarization of nucleons needs additional dedicated devices to be measured. 

The only previous determination of the $\vert G_{E}/G_{M} \vert$ and $\Delta\Phi$ for a baryon was performed at BESIII using the exclusive process $e^{+}e^{-} \to \Lambda \bar{\Lambda}$ at $\sqrt{s}$ = 2.396~GeV. The relative phase of the $\Lambda$ EMFFs was extracted by fitting the angular distributions~\cite{BESIII:2019nep}. Many theoretical activities~\cite{Yang:2019mzq, Haidenbauer:2020wyp, Mangoni:2021qmd, Dai:2021yqr, Lin:2022baj, Li:2021lvs} arose after this measurement. In Ref.~\cite{Haidenbauer:2020wyp}, the EMFF ratio and their relative phase are also predicted for $\Sigma$ hyperons, with a different dependence on the center-of-mass~(c.m.) energy from the $\Lambda$ case, reflecting complex dynamics. Though the $G_{\rm eff}$ and $\vert G_{E}/G_{M} \vert$ of the $\Sigma$ hyperons have been measured by various experiments~\cite{BESIII:2020uqk, BESIII:2021rkn, BaBar:2007fsu, Belle:2022dvb}, the extraction of $\Delta\Phi$ for $\Sigma$ is still unavailable. Thus, measurements of $\Sigma$ EMFFs can provide deeper insight into $\bar{Y}Y$ dynamics. Moreover, analyticity implies that the EMFFs tend to be real at large four-momentum transfer squared in the timelike region~\cite{Mangoni:2021qmd}. Since $\sin\Delta \Phi_{\Lambda}$ has previously been found to be significantly different from zero~\cite{BESIII:2019nep}, this indicates that the asymptotic threshold has not yet been reached for the $q^{2}$ so far studied. The phase measurement in a broader four-momentum transfer squared range is thus important to ascertain the asymptotic behavior of the hyperons and to investigate its dynamical mechanisms~\cite{Mangoni:2021qmd}. 

In this letter, we present a study of $e^{+}e^{-} \to \Sigma^{+} \bar{\Sigma}^{-}$ at three energy points, $\sqrt{s}$ = 2.3960, 2.6454, and 2.9000~GeV, with a total integrated luminosity of 239.84~pb$^{-1}$ collected with the Beijing Spectrometer~(BESIII) at the Beijing Electron Positron Collider~(BEPCII). The first energy point, 2.3960~GeV, is in close proximity to the production threshold for $\Sigma^{+}$ hyperon pairs~($2M_{\Sigma^{+}}=2.3788$~GeV), where $M_{\Sigma^{+}}$ represents the nominal mass of the $\Sigma^{+}$~\cite{Workman:2022ynf}. Here 2.6454~GeV is a combined dataset of 2.6444~GeV and 2.6464~GeV. The $\vert G_{E}/G_{M} \vert$ ratio and the relative phase $\Delta\Phi$ are determined using a fully differential angular expression. The formalism is described in Ref.~\cite{Perotti:2018wxm}. 

\begin{figure*}[htbp!]
\begin{center}
    \begin{overpic}[width=5.9cm,angle=0]{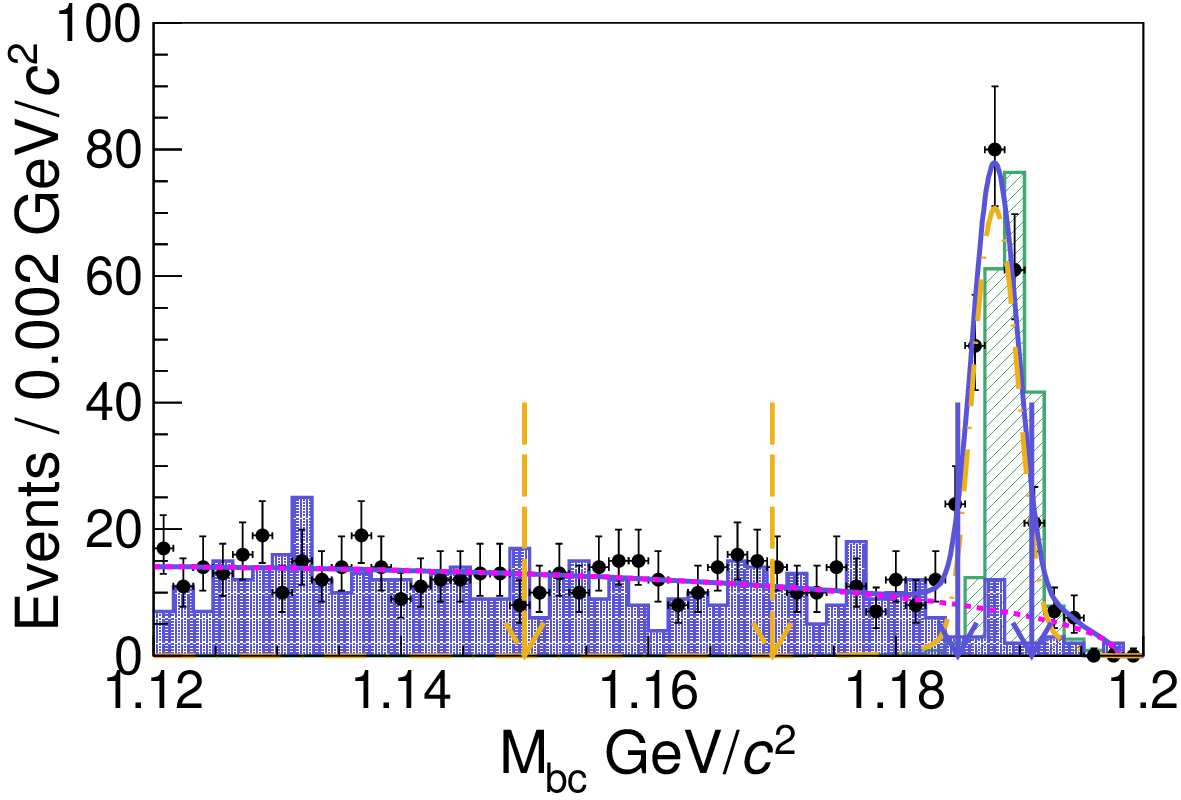}
    \put(17,58){\textbf{(a)}}
    \end{overpic}
    \begin{overpic}[width=5.9cm,angle=0]{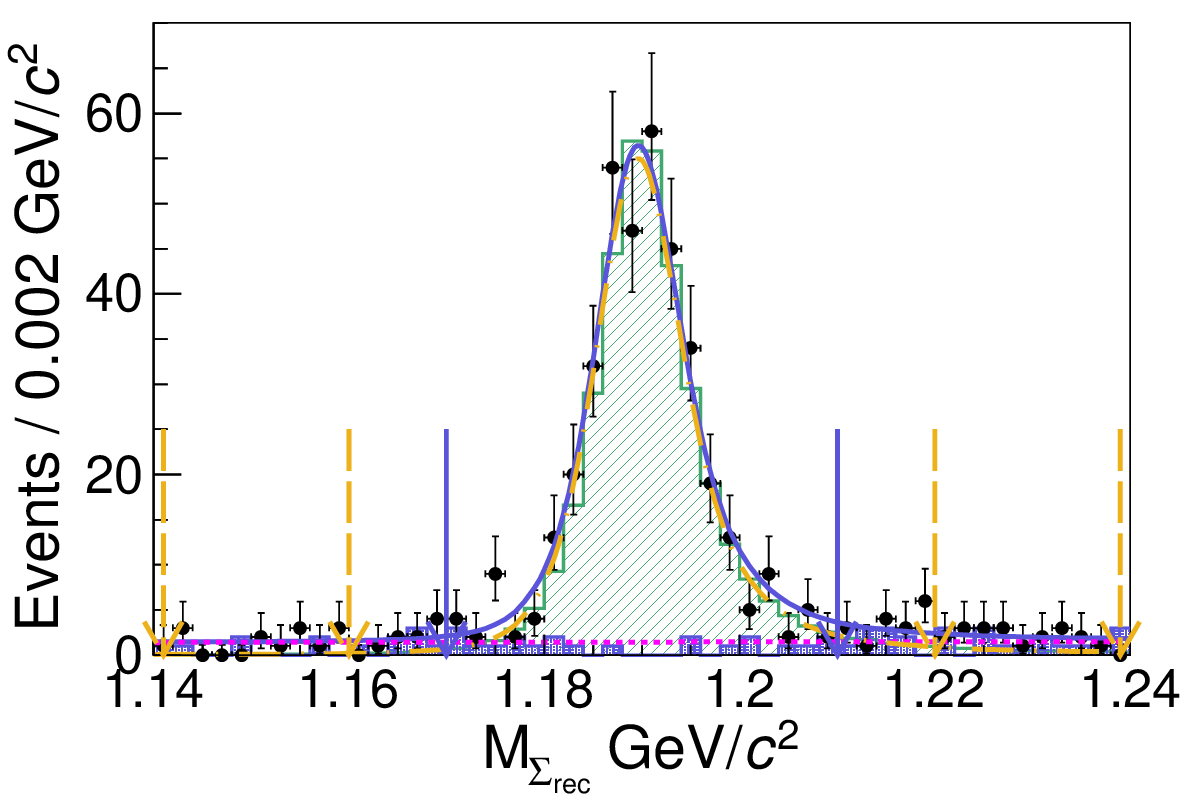}
    \put(17,58){\textbf{(b)}}
    \end{overpic}
    \begin{overpic}[width=5.9cm,angle=0]{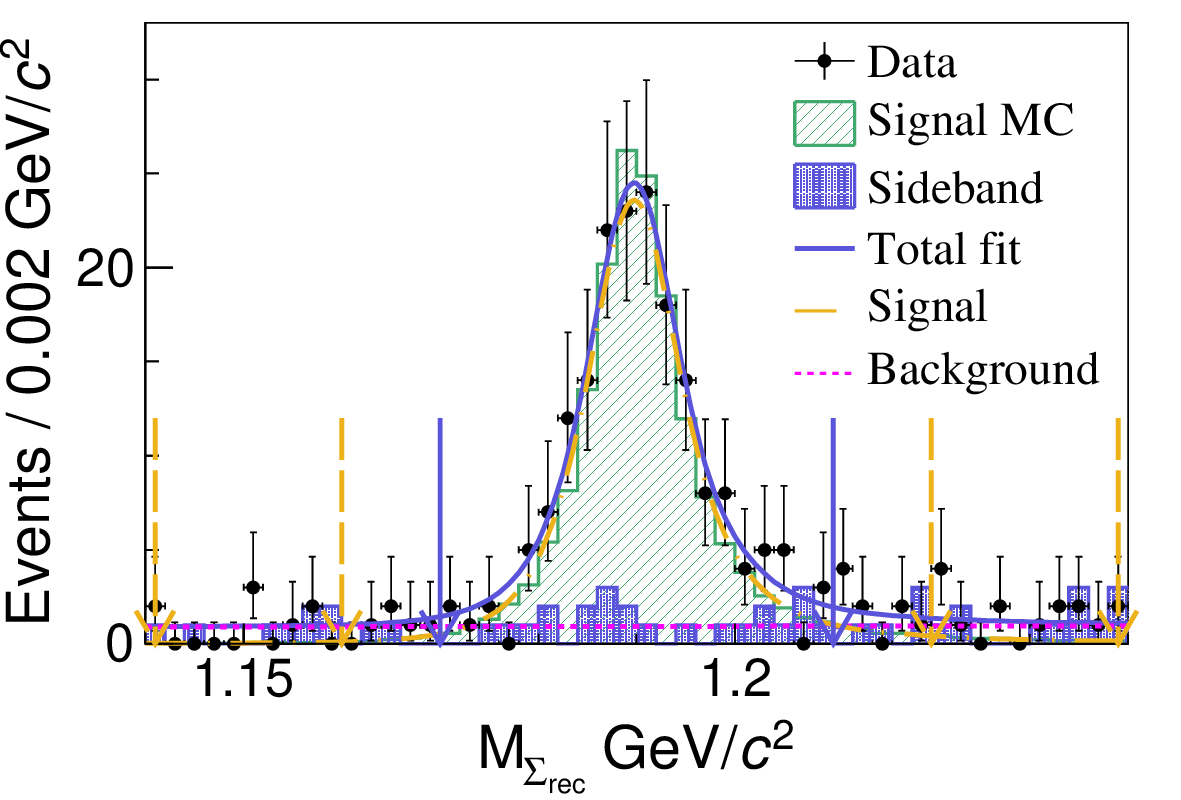}
    \put(17,58){\textbf{(c)}}
    \end{overpic}
    \end{center}
\caption{
The distributions of (a) $M_{\rm bc}$ at 2.3960~GeV, (b) $M_{\Sigma_{\rm rec}}$ at 2.6454~GeV, and (c) $M_{\Sigma_{\rm rec}}$ at 2.9000~GeV. The black dots with error bars are data. The histograms filled with green diagonal lines represent the signal MC samples, and the histograms filled with purple shading represent the backgrounds estimated by the sidebands. The purple solid lines are the total fit result. The yellow dash-dotted and magenta dotted lines are the signal and background shapes, respectively. The signal and background regions used for further angular analysis are indicated by purple solid-line arrows and yellow dashed-line arrows, respectively.}
\label{Fit}
\end{figure*}

The description of the design and performance of the BESIII detector can be found in Ref.~\cite{BESIII:2009fln}. The Monte Carlo~(MC) samples used to optimize event selection criteria are generated using a {\sc GEANT4}-based~\cite{GEANT4:2002zbu} simulation software package. The {\sc CONEXC}~\cite{Ping:2013jka} generator is used to generate signal MC samples and includes higher order processes with one radiative photon in the final state. The input cross section of line shape for $e^{+}e^{-} \to \Sigma^{+} \bar{\Sigma}^{-}$ is obtained from Ref.~\cite{BESIII:2020uqk}. The phase space~(PHSP) model in {\sc EvtGen}~\cite{Lange:2001uf, Ping:2008zz} is used to generate six million MC events to calculate the normalization factors in the multidimensional fits. The inclusive MC sample is generated with a {\sc HYBRID} generator~\cite{Ping:2016pms} for background analysis at each energy point. 

Two different reconstruction methods are used to select $\Sigma^{+}\bar{\Sigma}^{-}$ pairs, according to the c.m.~energy. At $\sqrt{s}$ = 2.3960~GeV, due to the low tracking efficiency for low-momentum tracks, a single-tag method is used to select the process $e^{+}e^{-}\rightarrow\Sigma^{+}\bar{\Sigma}^{-}\rightarrow \bar{p}\pi^{0}+X$, where $X$ denotes inclusive decays of the $\Sigma^{+}$. At higher c.m.~energies, both proton and antiproton are selected in the process $e^{+}e^{-}\rightarrow\Sigma^{+}\bar{\Sigma}^{-}$. To improve the detection efficiency, only one $\pi^0$ is reconstructed by two photons. 

Charged tracks are reconstructed in the main drift chamber~(MDC) as in Ref.~\cite{BESIII:2020fqg}. Combined information of the specific ionization energy loss~(d$E$/d$x$) in the MDC and the time of flight~(TOF) is used to calculate particle identification~(PID) probabilities for the pion, kaon, and proton hypotheses. The particle type with the highest probability is assigned for the track. At $\sqrt{s}$ = 2.3960~GeV, only the d$E$/d$x$ is used for PID since the charged tracks cannot reach the TOF detector due to low momenta. Photon candidates are reconstructed from clusters of energy deposited in the electromagnetic calorimeter~(EMC) as in Ref.~\cite{BESIII:2020fqg}. To reject showers from charged tracks, the angle between the shower direction and the track extrapolated to the EMC must be greater than 20 degrees in the single-tag reconstruction.

In the single-tag reconstruction at $\sqrt{s}$ = 2.3960~GeV, at least one good charged track, identified as an antiproton, is required. At least two good photons are required in each event. The $\bar{\Sigma}^{-}$ candidates are selected by looping over all possible $\bar{p}\gamma\gamma$ combinations. Two variables, $\Delta E$ and $M_{\rm bc}$, which reflect energy and momentum conservation, are used to select $\bar{\Sigma}^{-}$ candidates. Here $\Delta E\equiv E-E_{\rm beam}$ is the energy difference, where $E$ is the total measured energy of the $\bar{\Sigma}^{-}$ and $E_{\rm beam}$ is the beam energy, and $M_{\rm bc}\equiv\sqrt{E^{2}_{\rm beam}/c^{4}-P^{2}_{\bar{\Sigma}^{-}}/c^{2}}$ is the beam-constrained mass and $P$ is the magnitude of measured total momentum of the $\bar{\Sigma}^{-}$ candidate. Further selection criteria on the $\gamma\gamma$ invariant mass~($M_{\gamma\gamma}$) and $\Delta E$, 0.126 $< M_{\gamma\gamma} <$  0.139~GeV/$c^{2}$ and -0.013 $< \Delta E < $ 0.005~GeV, are applied. After the above selections, the distribution of $M_{\rm bc}$ at $\sqrt{s}$ = 2.3960~GeV is shown in Fig.~\ref{Fit}(a).

In the reconstruction with one missing $\pi^{0}$ at $\sqrt{s}$ = 2.6454 and 2.9000~GeV, a good event must have at least two good charged tracks identified to be one proton and one antiproton. At least two good photons are selected, and $\pi^{0}$ candidates are reconstructed from pairs of photons as in Ref.~\cite{BESIII:2020fqg}. At least one good $\pi^{0}$ candidate is required. To further remove potential background and improve the mass resolution, a two-constraint~(2C) kinematic fit under the $e^{+}e^{-}\to p\bar{p}\pi^{0}\pi^{0}$ hypothesis is performed. The fit requires total energy-momentum conservation, and the $\gamma\gamma$ invariant mass is constrained to the nominal $\pi^{0}$ mass, while the other $\pi^{0}_{\rm miss}$ is treated as a missing particle with free three-momentum. For events with more than one $\pi^{0}_{\gamma\gamma}$ candidate, by looping over the $\pi^{0}_{\gamma\gamma}$ candidates in the kinematic fit, the best $\pi^{0}_{\gamma\gamma}$ is selected with the minimum $\chi^{2}_{2C}$ which is further required to be less than 15. The $\pi^{0}_{\gamma\gamma}$ is then paired with either the proton or antiproton depending on which combination gives the minimum $\vert M_{(p\pi^{0}_{\gamma\gamma}/\bar{p}\pi^{0}_{\gamma\gamma})} - M_{\Sigma^{+}} \vert$ and the best combination is denoted as $\Sigma_{\rm tag}$. The signal region in the invariant mass of $\Sigma_{\rm tag}$ is chosen as 1.175 $< M_{\Sigma_{\rm tag}} <$ 1.200~GeV/$c^{2}$. The recoiling mass spectrum against $\Sigma_{\rm tag}$, $M_{\Sigma_{\rm rec}}$, after the previously described selections, is shown in Figs.~\ref{Fit}(b)(c).

Both the inclusive MC sample and the data sideband are used to study the potential background events. The main background, found in the inclusive MC sample, includes processes from $e^{+}e^{-}$ annihilation events with the same final states as the signal, with an additional photon, and with intermediate states like $\Lambda$, $\Sigma$ and $\Delta$ baryons. The background in the inclusive MC sample is smooth. The sideband regions are defined as -0.040 $<\Delta E<$ -0.031~GeV and 0.028 $<\Delta E<$ 0.037~GeV for $\sqrt{s}$ = 2.3960~GeV, and 1.135 $< M_{\Sigma_{\rm tag}} <$ 1.150~GeV/$c^{2}$ and 1.225 $< M_{\Sigma_{\rm tag}} <$ 1.240~GeV/$c^{2}$ for other energy points. As shown in Fig.~\ref{Fit}, the backgrounds in the sideband regions in both $M_{\rm bc}$ and $M_{\Sigma_{\rm rec}}$ are smooth, so no further selection is applied. 

To extract the signal yield, a simultaneus fit of $M_{\rm bc}$ and $M_{\Sigma_{\rm rec}}$ is applied. In the fit, the probability density functions~(PDF) of signal events are described by MC-simulated shapes, extracted from the signal MC sample, convolved with a Gaussian function. The PDFs of background events are described by an Argus function~\cite{ARGUS:1990hfq} at $\sqrt{s}$ = 2.3960~GeV and a linear function at $\sqrt{s}$ = 2.6454 and 2.9000~GeV. The best fit results are shown in Fig.~\ref{Fit}. The numbers of signal events are 207$\pm$17, 364$\pm$21, and 168$\pm$15 at 2.3960, 2.6454, and 2.9000~GeV, respectively, and the corresponding MC selection efficiencies are 11.33\%, 34.39\%, and 33.58\%, respectively. Furthermore, a cross-check of the Born cross section with the previous BESIII results~\cite{BESIII:2020uqk} is performed to ensure the reliability of the selection method. To ensure a pure sample for the further angular distribution analysis, tighter selections are applied on both $M_{\rm bc}$ and $M_{\Sigma_{\rm rec}}$, requiring 1.185 $< M_{\rm bc} <$ 1.191~GeV/$c^{2}$  and 1.170 $< M_{\Sigma_{\rm rec}} <$ 1.210~GeV/$c^{2}$ as indicated with arrows in Fig.~\ref{Fit}. The background fractions are 12.7$\%$, 7.7$\%$, and 10.2$\%$ at 2.3960, 2.6454, and 2.9000~GeV, respectively.

Following Refs.~\cite{Perotti:2018wxm, Faldt:2017kgy}, the joint angular distribution $\mathcal{W(\xi)}$ of $e^{+}e^{-} \to \Sigma^{+}~(\to p\pi^{0}) \bar{\Sigma}^{-}~(\to \bar{p}\pi^{0})$ can be expressed as
\begin{widetext}
\begin{equation}
\begin{aligned}
\label{angular distribution}
    \mathcal{W(\xi)} \propto &\mathcal{F}_{0}(\xi) + \alpha \mathcal{F}_{5}(\xi) + \alpha_{1}\alpha_{2}[\mathcal{F}_{1}(\xi) + \sqrt{1-\alpha^{2}}\cos(\Delta\Phi)\mathcal{F}_{2}(\xi) +\alpha\mathcal{F}_{6}(\xi)]\\& + \sqrt{1-\alpha^{2}}\sin(\Delta\Phi)[-\alpha_{1}\mathcal{F}_{3}(\xi) + \alpha_{2} \mathcal{F}_{4}(\xi)],
\end{aligned}
\end{equation}
\end{widetext}
where $\xi$ is a five-dimensional vector, $\xi$= $(\theta_{\Sigma^{+}}, \theta_{1}, \theta_{2}, \phi_{1}, \phi_{2})$; $\theta_{\Sigma^{+}}$ is the angle between the $\Sigma^{+}$ hyperon and positron beam; $\theta_{1}$~($\theta_{2}$) and $\phi_{1}$~($\phi_{2}$) are the polar and azimuthal angles of the proton~(antiproton) with respect to the $\Sigma^{+}$ and $\bar{\Sigma}^{-}$ helicity frame, respectively; and $\alpha_{1}$ and $\alpha_{2}$ are the decay asymmetry parameters of the $\Sigma^{+}$ and $\bar{\Sigma}^{-}$. The set of angular distribution functions $\mathcal{F}_{i}(\xi)$~($i$ = 0, 1, ..., 6) are obtained in Ref.~\cite{Perotti:2018wxm}. Owing to limited statistics, we assume $CP$ to be conserved and $\alpha_{1}$ = $-\alpha_{2}$ = $-0.980$~\cite{Workman:2022ynf}. The $\alpha$ is the angular distribution parameter describing the ratio of the two helicity amplitudes in $e^{+}e^{-}\to \Sigma^{+}\bar{\Sigma}^{-}$ and $\Delta\Phi$ is their relative phase. The $\alpha$ relates to $\vert G_{E}/G_{M} \vert$ via~\cite{Pacetti:2014jai} 
\begin{equation}
\begin{aligned}
\vert G_{E}/G_{M} \vert = \sqrt{\frac{s(1-\alpha)}{4M_{\Sigma^{+}}^{2}(1+\alpha)}}. 
\end{aligned}
\end{equation}
Since only one hyperon is reconstructed at $\sqrt{s}$ = 2.3960~GeV, $\theta_{1}$ and $\phi_{1}$ are integrated at this energy point, and the angular distribution becomes
\begin{equation}
\label{singleW}
    \begin{aligned}
    \mathcal{W(\xi)} & \propto \mathcal{F}_{0}(\xi) + \alpha \mathcal{F}_{5}(\xi)+\sqrt{1-\alpha^{2}}\sin(\Delta\Phi)\alpha_{2}\mathcal{F}_{4}(\xi).
    \end{aligned}
\end{equation}

The parameters $\alpha$ and $\Delta\Phi$ can be extracted by a multidimensional maximum likelihood fit to data. The joint likelihood function for observing $N$ events in the data sample is
\begin{equation}
    \begin{aligned}
    \mathcal{L}=\prod_{i=1}^{N}\mathcal{P}(\xi_{i};\alpha ,\Delta\Phi)=\prod_{i=1}^{N}\mathcal{C}\mathcal{W}(\xi_{i};\alpha ,\Delta\Phi)\epsilon(\xi_{i}), 
    \end{aligned}
\end{equation}
where $\mathcal{P}(\xi_{i};\alpha, \Delta\Phi)$ is the probability density function of $\xi_{i}$, $i$ is the corresponding event index, and $\epsilon(\xi_{i})$ is the efficiency of each event. The normalization factor $\mathcal{C}$ is given by $\mathcal{C}^{-1}=\int{\mathcal{W}(\xi;\alpha, \Delta\Phi)\epsilon(\xi)d\xi}$ and evaluated by the PHSP signal MC sample. The parameters $\alpha$ and $\Delta\Phi$ are extracted by minimizing the likelihood function 
\begin{equation}
\label{Lfunction}
    S = -{\rm ln}~\mathcal{L}_{\rm Data}+{\rm ln}~\mathcal{L}_{\rm Bkg}, 
\end{equation}
where $\mathcal{L}_{\rm Data}$ is the corresponding likelihood value of data and $\mathcal{L}_{\rm Bkg}$ represents the background, estimated with data events in the background region indicated in Fig.~\ref{Fit} and normalized to the signal region. The best fit results for $\alpha$, $\Delta\Phi$, and~(or) $\sin(\Delta\Phi)$ are summarized in Table~\ref{summary}, where only $\sin(\Delta\Phi)$ can be extracted at 2.3960~GeV due to the application of a single-tag method and the lack of sufficient angular distribution information. 

\begin{table*}[htbp!]
 \begin{center}                          
\caption{Fit results for $\alpha$, $\Delta\Phi~(^\circ)$, $\sin(\Delta\Phi)$, and $\vert G_{E}/G_{M} \vert$ at each energy point.}  
\label{summary}
\begin{ruledtabular}
\begin{tabular}{cccc}                   

$\sqrt{s}$~(GeV) & 2.3960 & 2.6454 & 2.9000 \\

\hline        

$\alpha$& -0.47~$\pm$~0.18~$\pm$~0.09 & 0.41~$\pm$~0.12~$\pm$~0.06 & 0.35~$\pm$~0.17~$\pm$~0.15 \\

$\Delta\Phi~(^\circ)$ & -42 $\pm$ 22 $\pm$ 14~(-138 $\pm$ 22 $\pm$ 14) & 55~$\pm$~19~$\pm$~14 & 78~$\pm$~22~$\pm$~9 \\

$\sin\Delta\Phi$& -0.67~$\pm$~0.29~$\pm$~0.18 &\\

$\vert G_{E} / G_{M} \vert$& 1.69~$\pm$~0.38~$\pm$~0.20  & 0.72~$\pm$~0.11~$\pm$~0.06 & 0.85~$\pm$~0.16~$\pm$~0.15 \\

\end{tabular}
\end{ruledtabular}
\end{center}                                                   
\end{table*} 

\begin{figure}[htbp!]
\begin{center}
    \begin{overpic}[width=4.2cm,angle=0]{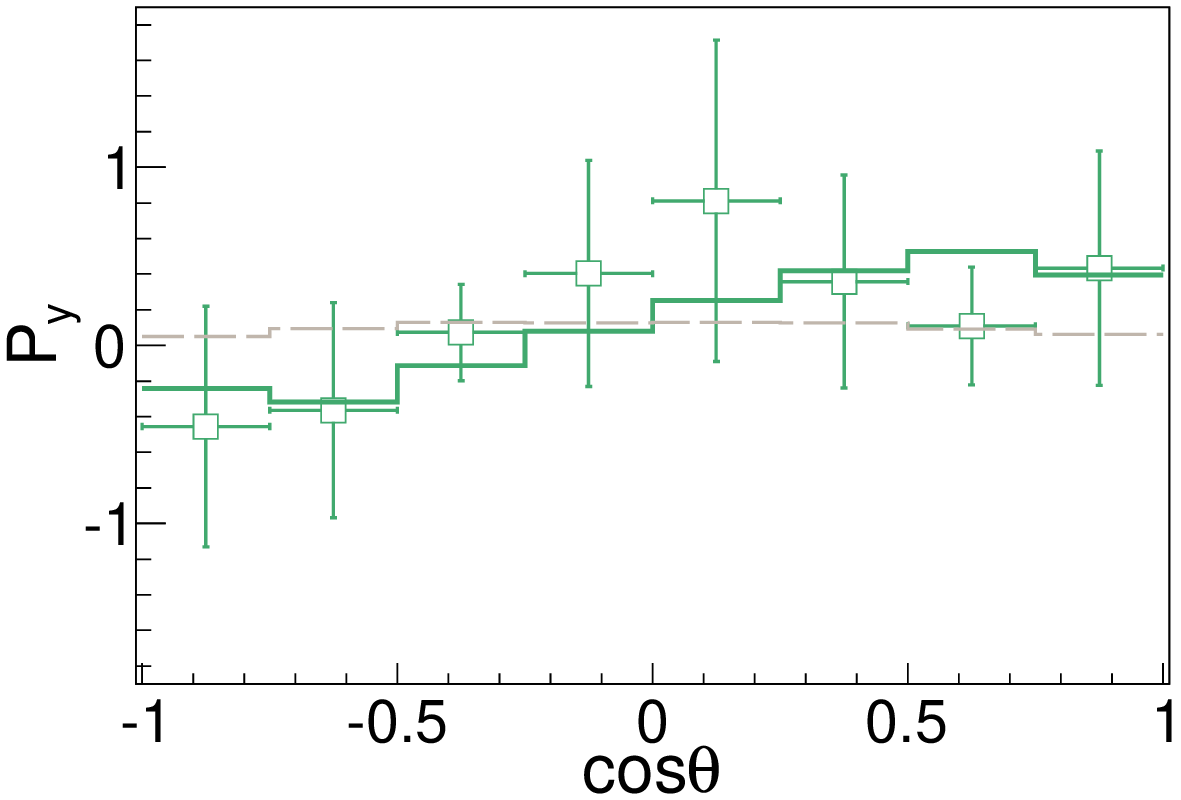}
    \put(17,58){\textbf{(a)}}
    \end{overpic}
    \begin{overpic}[width=4.2cm,angle=0]{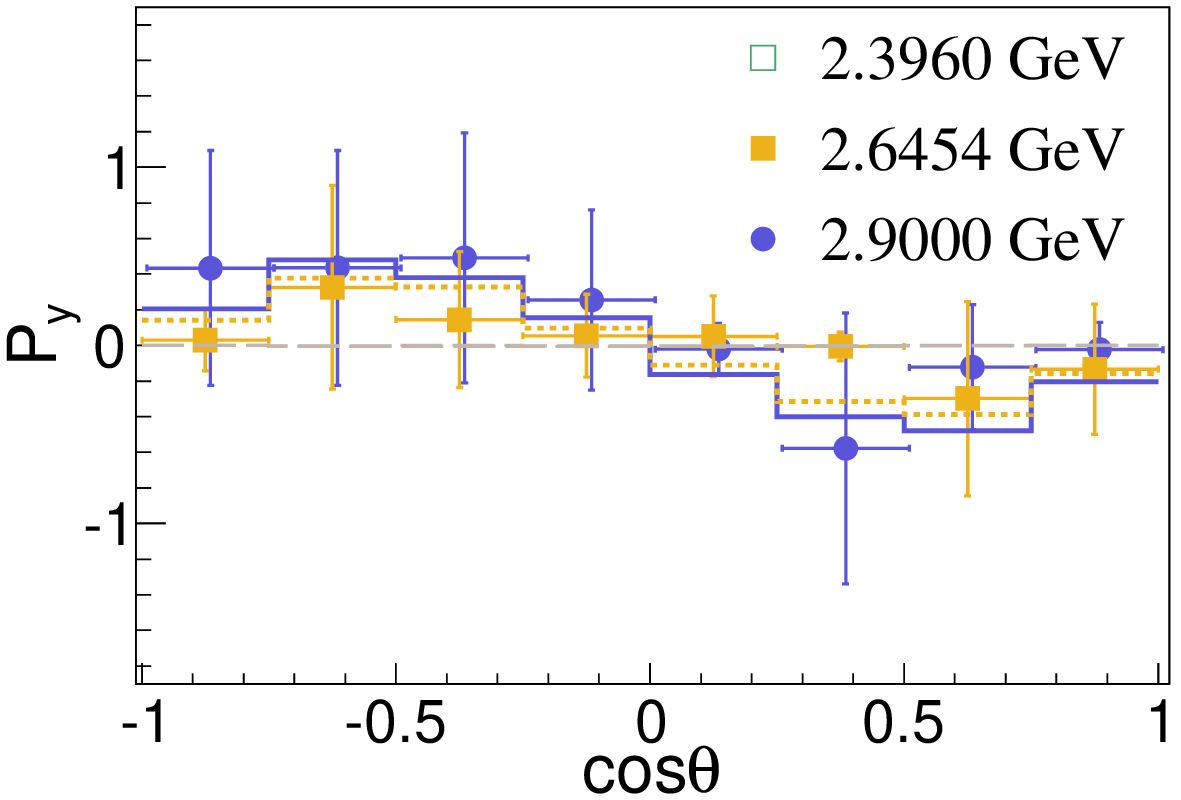}
    \put(17,58){\textbf{(b)}}
    \end{overpic}
\end{center}
\caption{The polarization $P_{y}$ as a function of the scattering angle at 2.3960~GeV~(a) and 2.6454 and 2.9000~GeV~(b). The open squares, solid squares, and dots are data. The histograms with solid lines~(dotted line at 2.6454~GeV) are signal MC samples based on the fit results, and the histograms with the gray dashed lines are the PHSP signal MC samples at each energy point.}
\label{moment}
\end{figure}

Furthermore, the nonzero $\Delta\Phi$ will lead to a dependence of the polarization on the scattering angle of the $\Sigma^{+}$~\cite{Dubnickova:1992ii, Faldt:2017kgy}:
\begin{equation}
    \begin{aligned}
	P_y= -\frac{\sqrt{1-\alpha^2}\sin\theta_{\Sigma^{+}}\cos\theta_{\Sigma^{+}}}{1+\alpha\cos^2\theta_{\Sigma^{+}}}\sin(\Delta\Phi). 
    \end{aligned}
	\label{pol_th}
\end{equation}
Experimentally, the $P_y$ is determined by
\begin{equation}
    \begin{aligned}
	P_y=\frac{m}{N}\sum_{i=1}^{N_{k}}\frac{(3+\alpha)(n_{1,y}^{i}+n_{2,y}^{i})}{(\alpha_{1}-\alpha_{2})(1+\alpha\cos^{2}\theta^{i}_{\Sigma^{+}})}, 
    \end{aligned}
	\label{pol_ex}
\end{equation}
where $N$ is the total number of events in the dataset and $m$ = 8 is the number of bins in $\cos\theta_{\Sigma^{+}}$; $N_{k}$ denotes the number of events in the $k$-th $\cos\theta_{\Sigma^{+}}$ bin; and $n_{1,y}$~($n_{2,y}$) is the projection of a proton~(antiproton) perpendicular to the scattering plane in the rest frame of $\Sigma^{+}$~($\bar{\Sigma}^{-}$). To test the goodness of the fit results, the signal MC sample is generated using Eqs.~(\ref{angular distribution}) and~(\ref{singleW}) and inputting the measured parameters from the data. The angular-dependent transverse polarization of $\Sigma$ is obtained as shown in Fig.~\ref{moment}.  

The sources of systematic uncertainties are summarized in Table~\ref{sys_sum}. For the first four sources in Table~\ref{sys_sum}, uncertainties are caused by the event selection and are evaluated by varying the selection criteria. For the fifth to eighth sources in Table~\ref{sys_sum}, the uncertainties from the fit procedure are estimated with alternative fits by varying the signal region, changing the sideband selections, and varying the fixed decay parameters~($\alpha_{1}$, $\alpha_{2}$) by $\pm$1$\sigma$, individually. The maximum difference with the nominal value is taken as the uncertainty. To estimate the systematic uncertainty of the fit method, 500 sets of signal MC samples with the parameters from Table~\ref{summary} are generated and fitted to obtain the distribution of the output parameters, and the difference between the input and averaged output values is assigned as the systematic uncertainty. Some inconsistencies between data and MC simulation are observed in the $M_{\rm bc}$ distribution, as shown in Fig.~\ref{Fit}(a). To estimate their effect on the final results, the measurement of beam energy and the calibration of the $\bar{\Sigma}^{-}$ momentum are investigated. For the $E_{\rm beam}$ calibration, we generate three MC samples with different c.m.~energies, defined around 2.3960~GeV in steps of 1~MeV, that is, 2.3950, 2.3970, and 2.3980~GeV, and choose the one that gives the best description of the data in the fit procedure. For the $\bar{\Sigma}^{-}$ momentum calibration, ten MC samples are generated, with different scale factors for the three-momentum of antiproton in each sample. The scale factors are defined in steps of 0.001 from 1.040 to 1.049, and we choose the one giving the best description of the data in the fit procedure. The differences between the updated and nominal results are taken as the systematic uncertainties. In Table~\ref{sys_sum}, the individual uncertainties are assumed to be uncorrelated and are added in quadrature.

\begin{table}[htbp!] 
    \begin{center} 
    \caption{The systematic uncertainties for $\alpha$, $\Delta\Phi$~$(^\circ)$, and $\sin(\Delta\Phi)$ at each energy point~(in GeV).} 
    \label{sys_sum}
     \begin{ruledtabular}
    \begin{tabular}{ccccccc}             
    \multirow{2}{*}{Source}  & \multicolumn{2}{c}{2.3960}  &   \multicolumn{2}{c}{2.6454}    &        \multicolumn{2}{c}{2.9000}   \\ 
    \cline{2-7} 
                        & $\alpha$  & $\sin(\Delta\Phi)$ & $\alpha$  & $\Delta\Phi$ &  $\alpha$  & $\Delta\Phi$ \\
    \hline
    $\Delta E$ cut        & 0.03      & 0.02               &           &                       &            &        \\
    $\gamma\gamma$ mass window  & 0.04      & 0.06               &           &                       &            &        \\
    $\chi^{2}_{2C}$ cut &           &                    & 0.04      & 5                  &  0.08      & 5   \\
    $\Sigma_{\rm tag}$ mass window  &           &                    & 0.00      & 3                  &  0.06      & 2   \\
    Signal region       & 0.05      & 0.16               & 0.04      & 9                 &  0.05      & 4  \\
    Sideband region    & 0.02      & 0.06               & 0.02      & 9                  &  0.09      & 5  \\
    $\alpha_{1}$        &           &                    & 0.01      & 0                 &  0.00      & 1  \\
    $\alpha_{2}$        & 0.00      & 0.01               & 0.01      & 0                  &  0.00      & 1  \\
    Fit method      & 0.00      & 0.01               & 0.02      & 2                  &  0.03      & 2    \\
    $E_{\rm beam}$ calibration & 0.03      & 0.00               &           &                       &            &\\
    Momentum calibration & 0.04      & 0.01               &           &    &   & \\
    \hline
    Total                 & 0.09      & 0.18               & 0.06      & 14                 &  0.15      & 9   \\                                            
    \end{tabular}                                                     
    \end{ruledtabular}
    \end{center}                                                       
\end{table}  

\begin{figure*}[htbp!]
\begin{center}
    \begin{overpic}[width=7.0cm,angle=0]{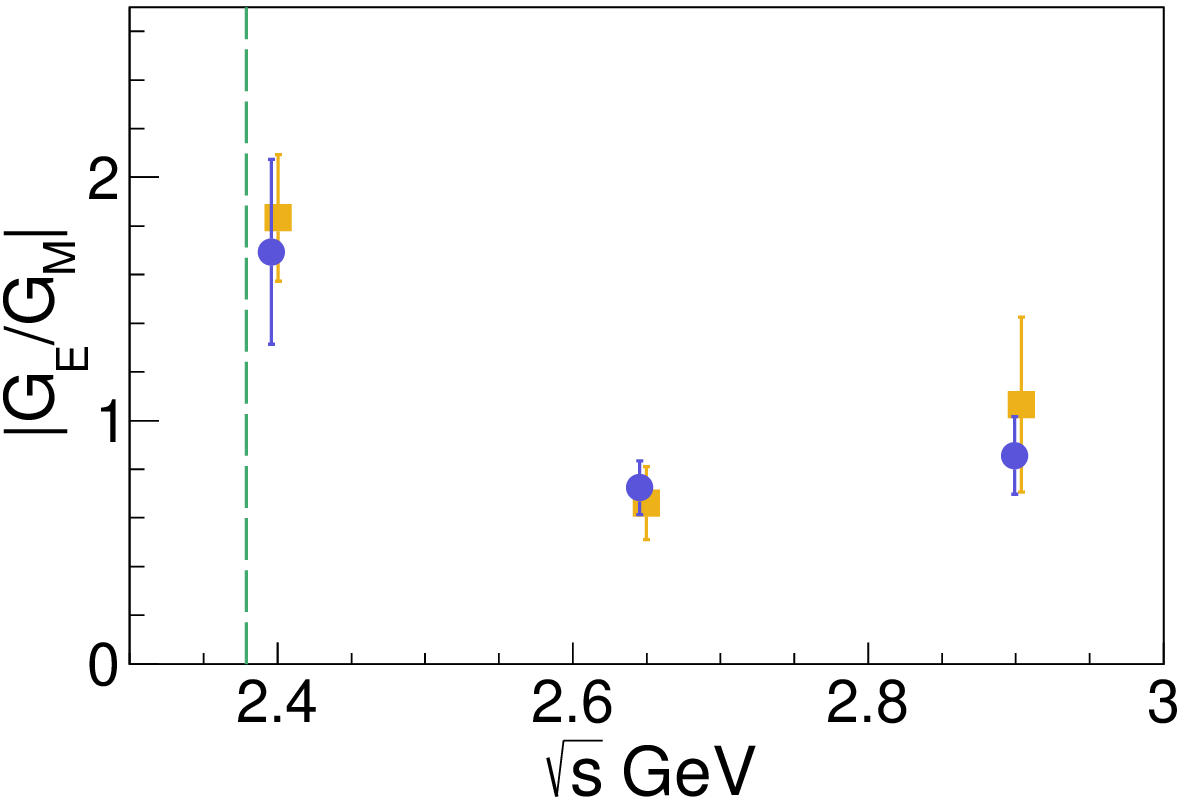}
    \put(23,60){\textbf{(a)}}
    \end{overpic}\hspace{1cm}
    \begin{overpic}[width=7.0cm,angle=0]{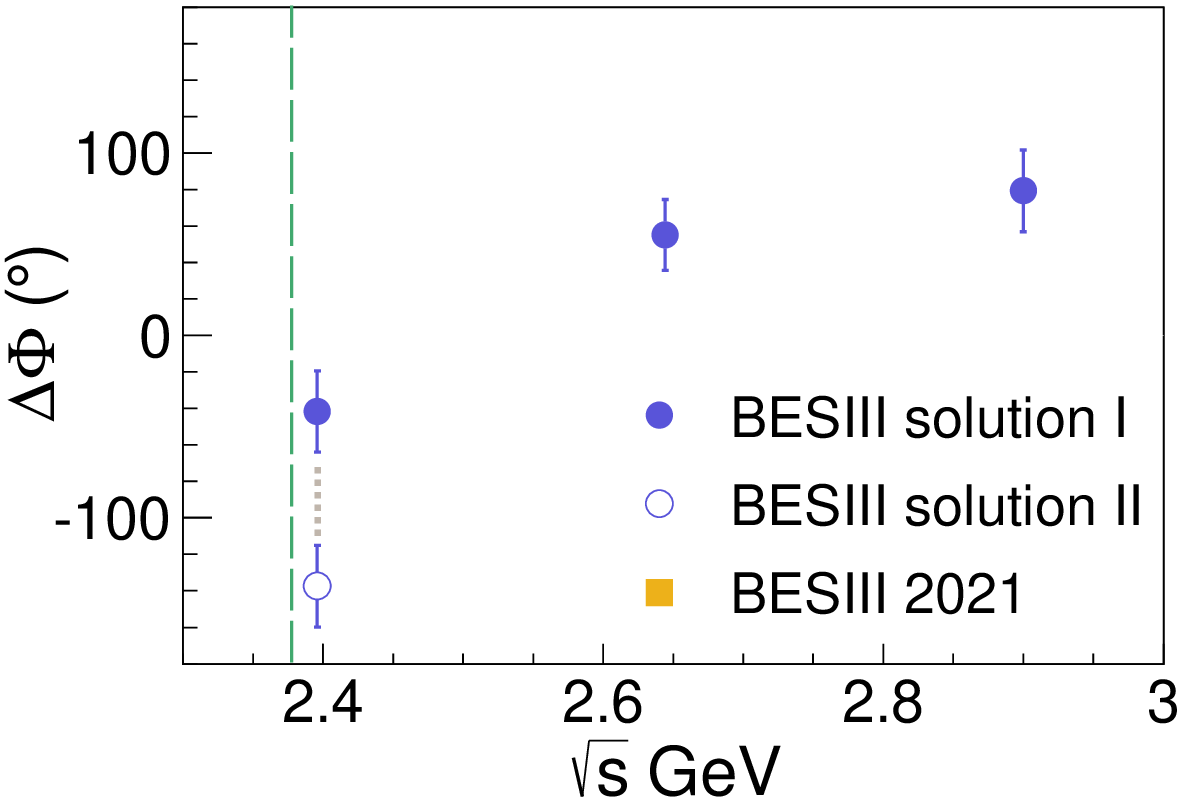}
    \put(27,60){\textbf{(b)}}
    \end{overpic}
\caption{Results for $\vert G_{E}/G_{M} \vert$~(a) and the relative phase $\Delta\Phi$~(b) from this work~(purple dots). The yellow squares in~(a) denote the previous results from BESIII~\cite{BESIII:2020uqk}. The open circle in~(b) represents the second solution of $\Delta\Phi$ at 2.3960~GeV. The vertical dashed lines indicate the production threshold for $e^{+}e^{-} \to \Sigma^{+}\bar{\Sigma}^{-}$, where $\vert G_{E}/G_{M} \vert$ = 1 and $\Delta\Phi$ = 0$^{\circ}$ by definition.}
\label{sum}
\end{center}  
\end{figure*}

In summary, the process $e^{+}e^{-} \to \Sigma^{+} \bar{\Sigma}^{-}$ is studied at 2.3960, 2.6454, and 2.9000~GeV. Using a joint angular distribution analysis, 
the final results for $\vert G_{E}/G_{M} \vert$, the relative phase $\Delta\Phi$, and $\sin\Delta\Phi$ are summarized in Table~\ref{summary} and plotted in Fig.~\ref{sum}, where the relative phase of the $\Sigma^{+}$ hyperon is measured for the first time in a wide four-momentum transfer range. 

The precision of $\vert G_{E}/G_{M} \vert$ is improved compared with the previous measurement~\cite{BESIII:2020uqk} at 2.6454 and 2.9000~GeV. Since only the sine value of $\Delta\Phi$ can be extracted at 2.3960~GeV, the two solutions are plotted as shown in Fig.~\ref{sum}(b), and there is a significant discrepancy between our experimental result for $\Delta\Phi$ and the theoretical predictions from the $\bar{Y}Y$ potential model~\cite{Haidenbauer:2020wyp}. On the other hand, in Fig.~\ref{sum}(b), $\Delta\Phi$ is less than zero at 2.3960~GeV and greater than zero at 2.6454~GeV, which implies that there may be at least one $\Delta\Phi=0^{\circ}$ between these two energy points. Such an evolution will be an important input for understanding its asymptotic behavior~\cite{Mangoni:2021qmd} and the dynamics of baryons. Moreover, the fact that the relative phase is still increasing at 2.9000~GeV indicates that the asymptotic threshold has not yet been reached.

\input{acknowledgement_2023-05-14.tex}
\bibliographystyle{apsrev4-1}
\bibliography{reference.bib}

\end{document}

%% file: authorlist_2023-05-14.tex
\author{
	\begin{small}
		\begin{center}
			M.~Ablikim$^{1}$, M.~N.~Achasov$^{5,b}$, P.~Adlarson$^{75}$, X.~C.~Ai$^{81}$, R.~Aliberti$^{36}$, A.~Amoroso$^{74A,74C}$, M.~R.~An$^{40}$, Q.~An$^{71,58}$, Y.~Bai$^{57}$, O.~Bakina$^{37}$
% ...
			\\
			\vspace{0.2cm}
			(BESIII Collaboration)\\
			\vspace{0.2cm} {\it
				$^{1}$ Institute of High Energy Physics, Beijing 100049, People's Republic of China\\
% ...
$^{5}$ Budker Institute of Nuclear Physics SB RAS (BINP), Novosibirsk 630090, Russia\\
% ...
$^{36}$ Johannes Gutenberg University of Mainz, Johann-Joachim-Becher-Weg 45, D-55099 Mainz, Germany\\
				$^{37}$ Joint Institute for Nuclear Research, 141980 Dubna, Moscow region, Russia\\
% ...
$^{40}$ Liaoning Normal University, Dalian 116029, People's Republic of China\\
% ...
$^{57}$ Southeast University, Nanjing 211100, People's Republic of China\\
				$^{58}$ State Key Laboratory of Particle Detection and Electronics, Beijing 100049, Hefei 230026, People's Republic of China\\
% ...
$^{71}$ University of Science and Technology of China, Hefei 230026, People's Republic of China\\
% ...
$^{74}$ University of Turin and INFN, (A)University of Turin, I-10125, Turin, Italy; (B)University of Eastern Piedmont, I-15121, Alessandria, Italy; (C)INFN, I-10125, Turin, Italy\\
				$^{75}$ Uppsala University, Box 516, SE-75120 Uppsala, Sweden\\
% ...
$^{81}$ Zhengzhou University, Zhengzhou 450001, People's Republic of China\\
				\vspace{0.2cm}
% ...
$^{b}$ Also at the Novosibirsk State University, Novosibirsk, 630090, Russia\\
% ...
		}\end{center}	
		\vspace{0.4cm}
	\end{small}
}

%% file: acknowledgement_2023-05-14.tex
The authors thank Professor L. Y. Dai for helpful discussion. The BESIII Collaboration thanks the staff of BEPCII and the IHEP computing center and the supercomputing center of USTC for their strong support. This work is supported in part by National Key R\&D Program of China under Contracts No. 2020YFA0406400, No. 2020YFA0406300 and No. 2023YFA1609400; National Natural Science Foundation of China (NSFC) under Contracts 
No. 11905092, No. 12105132, No. 11705078, No. 11625523, No. 12105276, No. 12122509, No. 11635010, No. 11735014, No. 11835012, No. 11935015, No. 11935016, No. 11935018, No. 11961141012, No. 12022510, No. 12025502, No. 12035009, No. 12035013, No. 12061131003, No. 12192260, No. 12192261, No. 12192262, No. 12192263, No. 12192264, No. 12192265, No. 12221005, No. 12225509, No. 12235017; the Chinese Academy of Sciences (CAS) Large-Scale Scientific Facility Program; the CAS Center for Excellence in Particle Physics (CCEPP); Joint Large-Scale Scientific Facility Funds of the NSFC and CAS under Contract No. U1732263, No. U1832103, No. U1832207 and No. U2032111; CAS Key Research Program of Frontier Sciences under Contracts No. QYZDJ-SSW-SLH003 and No. QYZDJ-SSW-SLH040; 100 Talents Program of CAS; The Institute of Nuclear and Particle Physics (INPAC) and Shanghai Key Laboratory for Particle Physics and Cosmology; The Double First-Class university project foundation of USTC; ERC under Contract No. 758462; European Union's Horizon 2020 research and innovation programme under Marie Sklodowska-Curie grant agreement under Contract No. 894790; German Research Foundation DFG under Contracts No. 455635585, Collaborative Research Center CRC 1044, FOR5327, GRK 2149; Istituto Nazionale di Fisica Nucleare, Italy; Ministry of Development of Turkey under Contract No. DPT2006K-120470; National Research Foundation of Korea under Contract No. NRF-2022R1A2C1092335; National Science and Technology fund of Mongolia; National Science Research and Innovation Fund (NSRF) via the Program Management Unit for Human Resources \& Institutional Development, Research and Innovation of Thailand under Contract No. B16F640076; Polish National Science Centre under Contract No. 2019/35/O/ST2/02907; The Swedish Research Council; U. S. Department of Energy under Contract No. DE-FG02-05ER41374; The PhD Start-up Fund of Natural Science Foundation of Liaoning Province of China under Contract No. 2019-BS-113; Education Department of Liaoning Province Scientific research Foundation of Liaoning Provincial Department of Education under Contracts No. LQN201902; Foundation of Innovation team 2020, Liaoning Province; Opening Foundation of Songshan Lake Materials Laboratory, Grants No.2021SLABFK04. CAS Youth Team Program under Contract No. YSBR-101. Knut \& Alice Wallenberg Foundation, Contract No. 2021.0174 and No. 2021.0299; Swedish Research Council, Contract No. 2019.04594; The Swedish Foundation for International Cooperation in Research and Higher Education, CH2018-7756.